\documentclass[conference, onecolumn]{IEEEtran}
\IEEEoverridecommandlockouts
\usepackage{cite}
\usepackage{amsmath,amssymb,amsfonts}
\usepackage{algorithmic}
\usepackage{graphicx}
\usepackage{textcomp}
\usepackage{xcolor}
\def\BibTeX{{\rm B\kern-.05em{\sc i\kern-.025em b}\kern-.08em
    T\kern-.1667em\lower.7ex\hbox{E}\kern-.125emX}}
    
\newcommand{\method}{\textsc{Phased}\xspace}
\usepackage{calc,bm,color,theorem,graphicx, dsfont}
\usepackage{hyperref} 
\usepackage{url}
\usepackage{MnSymbol}
\usepackage{algorithm}
\newcommand{\R}{\mathbb{R}}
\newcommand{\setV}{\mathcal{V}}
\newcommand{\setA}{\mathcal{A}}
\newcommand{\setT}{\mathcal{T}}
\newcommand{\setI}{\mathcal{I}}
\newcommand{\setS}{\mathcal{S}}
\newcommand{\setX}{\mathcal{X}}
\newcommand{\setY}{\mathcal{Y}}
\newcommand{\mathbbm}[1]{\text{\usefont{U}{bbm}{m}{n}#1}}

\usepackage{multirow}
\usepackage{xspace}
\usepackage{subfig}

\newtheorem{lemma}{Lemma}

\newtheorem{proposition}{Proposition}

\begin{document}

\title{\method: Phase-Aware Submodularity-Based Energy Disaggregation\\
\thanks{
The work of A. Konar and N. D. Sidiropoulos was supported in part by the National Science Foundation under Grant NSF IIS-1908070.
This work was authored in part by the National Renewable Energy Laboratory, operated by Alliance for Sustainable Energy, LLC, for the U.S. Department of Energy (DOE) under Contract No. DE-AC36-08GO28308. The work of A. S. Zamzam was supported in part by the Laboratory Directed Research and Development Program at the National Renewable Energy Laboratory.  The views expressed in the article do not necessarily represent the views of the DOE or the U.S. Government. The U.S. Government retains and the publisher, by accepting the article for publication, acknowledges that the U.S. Government retains a nonexclusive, paid-up, irrevocable, worldwide license to publish or reproduce the published form of this work, or allow others to do so, for U.S. Government purposes.

}
}

\author{\IEEEauthorblockN{Faisal M. Almutairi$^{\star}$, Aritra Konar$^{\dagger}$, Ahmed S. Zamzam$^{\mp}$, and Nicholas D. Sidiropoulos$^{\dagger}$}
\IEEEauthorblockA{{$\star$ ECE Dept., University of Minnesota, Minneapolis, MN} \\
\IEEEauthorblockA{{$^{\dagger}$ ECE Dept., University of Virginia, Charlottesville, VA} 
\IEEEauthorblockA{$^{\mp}$ National Renewable Energy Laboratory, Golden, CO}
\\
}}
}

\bibliographystyle{IEEEtran}
\maketitle

\begin{abstract}
Energy disaggregation is the task of discerning the energy consumption of individual appliances from aggregated measurements, which 
holds promise for understanding and reducing energy usage.
In this paper, we propose \method, an optimization approach for energy disaggregation that has two key features: \method(i) exploits the structure of power distribution systems to make use of readily available measurements that are neglected by existing methods, and (ii) poses the problem as a minimization of a difference of submodular functions. We leverage this form by applying a discrete optimization variant of the majorization-minimization algorithm to iteratively minimize a sequence of global upper bounds of the cost function to obtain high-quality approximate solutions.          
\method improves the disaggregation accuracy of state-of-the-art models by up to $61\%$ and achieves better prediction on heavy load appliances. 
\end{abstract}


\section{Introduction}
\label{sec:intro}
Improving the energy efficiency of smart homes via machine learning (ML) methods constitutes an important research area with many potential benefits, such as reducing the adverse effects of energy consumption on the environment. Energy disaggregation/non-intrusive load monitoring (NILM) seeks to
break down the energy usage of multiple household appliances from a single aggregated power measurement \cite{hart1992nonintrusive}. 
{NILM benefits a plethora of applications in the areas of energy saving, automation in smart homes, anomaly detection, and life coaching and recommendations \cite{shin2019data}.}     

Many ML approaches have been proposed for NILM; see \cite{faustine2017survey} and the references therein. Because the problem can be very ill posed, these methods are primarily supervised and require appliance-level training data available from homes with submeters (e.g., {data summarized in} \cite{parson2015dataport}) for learning a model that generalizes to new (unseen) homes using only their aggregated power consumption.
In this direction, sparse coding \cite{kolter2010energy,Powerlets,pandey2019structured} and matrix/tensor factorization \cite{rahimpour2017non,batra2018transferring,zamzam2020grate} approaches aim to learn a latent factor/dictionary from a training set, which is then used for disaggregation. 
The work in \cite{kolter2010energy} proposed a customized dictionary learning method, where appliance-specific bases are learned from labeled training data such that the disaggregation error is minimized. 
Another approach in \cite{rahimpour2017non} used nonnegative matrix factorization (NMF), where one factor corresponds to the normalized appliance-level power consumption as the basis. The other factor forms the basis coefficients, which are constrained to add up to 1 for each appliance to impose the ``groupin'' effect. 
Although they are conceptually appealing, these methods require large training data to capture all possible appliance states, and they depend on the (hard to validate) assumption of common latent factors between the training and test sets.
Neural network models have been deployed for the NILM task \cite{kelly2015neural, zhang2018sequence}.
For instance, the work in \cite{zhang2018sequence} proposed a network architecture, called sequence-to-point (seq2p), where the input is a window of the aggregated time series, and the output corresponds to the appliance power at the middle point in the given window.
In addition to its large number of trainable parameters ($>$ 30 M), the main drawback of seq2p is that it trains a separate model for each appliance independently; thus, it ignores the dependency among appliances (the aggregated signal is a joint function of all the constituent appliances). 


{Recently, the work in \cite{ouricassp} demonstrated that the energy disaggregation can be posed as a constrained set-function maximization problem, which is NP–hard in its general form. 
The authors proposed a discrete block successive approximation algorithm that exploits the fact that the cost function is block-submodular \cite{ouricassp}. Building on this line of work,
we propose \method\footnote{Code is available at: \url{https://github.com/FaisalAlmutairi/submodularity_based_NILM}.}, a supervised framework for energy disaggregation that leverages the connectivity structure of the power distribution networks. 
To the best of our knowledge, there has not been \emph{any} preexisting method that exploits such information. This allows us to obtain {\em multiple} aggregated measurements for each time instant, instead of a single measurement, thereby reducing the under-determinacy of the problem. 
Using appliance-level training data, we first learn to which energized line (phase) an appliance is connected and the appliance consumption levels at its different states (`on', `off', `standby', etc.). The effectiveness of this model in breaking down aggregated signals is then evaluated on the test set.
Although this requires solving a challenging, NP--hard, combinatorial optimization problem, we prove that the cost function can be decomposed as a \emph{difference of submodular functions} (DSF).
Leveraging the special properties of submodular functions \cite{fujishige2005submodular,bach2013learning}, we devise an efficient successive approximation algorithm for computing high-quality, albeit suboptimal solutions for the problem.
In contrast to \cite{ouricassp}, we establish
that the cost function can be expressed in DSF form
over the \emph{entire} time horizon, which results in a discrete approximation algorithm that features more attractive ``all-at-once'' updates. 
\method improves the error of four distinct classes of state-of-the-art 
approaches by up to $61\%$ when averaged over appliances.     
}


\begin{figure}[t]
	\centering
	{\includegraphics[width=5cm,height=2.5cm]{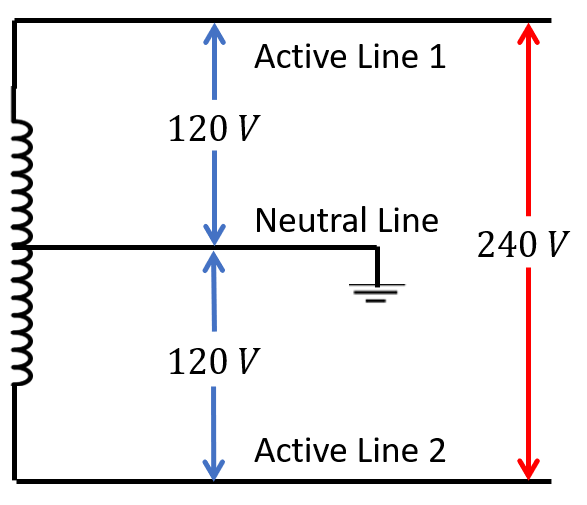}\label{a}}
	\hspace{5pt} 
	{\includegraphics[width=5cm,height=2.5cm]{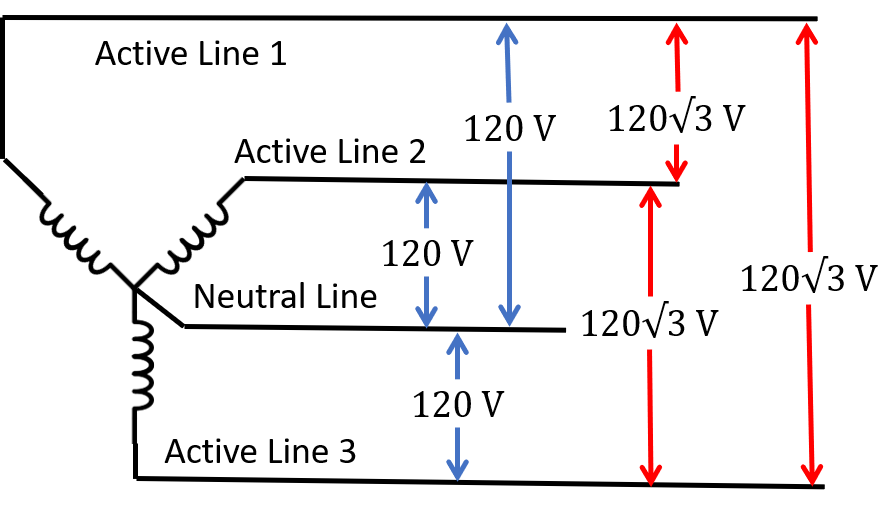}}
	\setlength{\belowcaptionskip}{-8pt}
	\caption{Split-phase (left) and three-phase (right).}
	\label{fig:1}
\end{figure}
\section{Problem Statement}
Given a household outfitted with $L$ appliances, let $\{y_t\}_{t=1}^T$ represent the time series of the aggregated power consumption. 
The goal of energy disaggregation is to 
decompose $y_t$ into $L$ components of the form $y_t = \sum_{i=1}^{L} x_{i,t}$, where $x_{i,t}$
denotes the power consumption of appliance $i$ at time $t$.
A particularly challenging aspect of the problem is that it can be very under-determined because  we wish to infer the power consumption of multiple appliances from a single measurement;
however, in practice, the power distribution system supplying a household with electricity comprises multiple power lines, each corresponding to a different phase. The aggregated power consumption at a given instant, $t$, then comprises multiple measurements, $\{y^{r}_t\}_{r=1}^R$, where $y^r_t$ is the power measured at the $r^{th}$ line (wire) at time $t$, and $R\in\{2, 3\}$ is the number of lines depending on the low-voltage connection.
The electrical networks usually employ one of the following two connections: 

\noindent (i)~\textbf{single-phase:}
also known as {\em split-phase}, commonly used in North America for residential buildings. 
In this connection, the transformer takes a single-phase input and provides a 240-V output with a center tap that is connected to the ground, i.e., it provides 240-V that is divided into two 120-V live conductors. 
Light loads are connected between a live conductor and the neutral to receive 120-V, whereas heavy loads receive 240-V by being connected between two live conductors \cite{elnozahy2013comprehensive}---see Figure~1.

\noindent (ii)~\textbf{three-phase:}
common in commercial buildings in the United States and in residential buildings in Europe. In this connection, the power is delivered over three live conductors. 
The premises are fed with four lines (three live conductors and a neutral) \cite{wildi2002electrical}. Each live conductor corresponds to a single phase with a phase separation of $120^{\circ}$ between any two live conductors. 
Figure~1 (right) shows the three-phase wye connection.

For both single- and three-phase-connected buildings, power consumption readings are often taken at every live conductor; 
however, the prevailing approaches in the literature do not consider the connectivity structure of the electrical feeder, and they make the simplifying assumption that aggregated power at a given time instant is drawn only from a single line, i.e., $R = 1$. Consequently, available information from other lines is summed, or even neglected. 

\section{Overview of Submodular Functions}
Given a ground set of $n$ elements, $\mathcal{V} := \{v_1, \cdots, v_n\}$, consider the set function $f:2^{\setV} \rightarrow \R$ that assigns a real value to any subset $\mathcal{S} \subseteq \mathcal{V}$. Among the set functions, the subclass of the \emph{submodular} functions is notable for exhibiting many properties similar to both convex and concave functions, and it arises in many applications in machine learning \cite{bach2013learning}. Formally, a set function, $f(.)$, is submodular if and only if it satisfies $f(\mathcal{X} \cup \{v\})-f(\mathcal{X}) \geq f(\mathcal{Y} \cup \{v\})-f(\mathcal{Y})$ for all $\mathcal{X}\subseteq\mathcal{Y}\subseteq \mathcal{V} \backslash 
\{v\}$. 	
That is, given any subset of elements $\setX$, the marginal gain derived by adding an element $v$ to $\setX$ does not increase when we instead add $v$ to the superset $\setY$. Hence, submodular functions exhibit a natural diminishing returns property. 
A submodular set function $y(.)$ is said to be \emph{modular} if and only if there exists a vector ${\bf y} \in \mathbb{R}^n$ for all subsets $\mathcal{X} \subseteq \mathcal{V}$ such that $y(\mathcal{X}) = {\bf y} ^T\mathbbm{1}_{\mathcal{X}} = \sum_{e \in \mathcal{X}} {\bf y} (e)$.

\section{Proposed Method: \method}
Our proposed method, \method, is cognizant of the underlying residential feeder topology and exploits the readily available multiple aggregated power measurements---each corresponding to the power drawn from one of the lines supplying the household---to reduce the under-determinacy of the problem. Note that 
a particular appliance can be connected between either one of the live lines and the neutral, or between two live lines.
Consequently, appliances that are connected to only one live line draw all of their consumed power from this particular line, whereas appliances connected between two live lines draw power from both lines. Formally:
\begin{equation}\label{eq1}
\begin{aligned}
& y^{r}_t = 
\textstyle \sum_{i=1}^{L} {w}_i^{r} {x}_{i,t},\;\forall \;r\in [R],
\end{aligned}
\end{equation}
\noindent If appliance $i$ is connected to only a single line $r \in [R]$ (and the neutral), then $w_i^r=1$ and $w_i^s=0$, $\forall\; s \neq r$. Otherwise, if appliance $i$ is connected between a pair of lines $(r,s)$, then $0<w_i^r<1$, $0<w_i^s<1$, and $w_i^r+w_i^s=1$.

We make the standard assumption that the power consumption profile of every appliance $i$ can be approximated by a finite number $N_i \geq 2$ of {\em states} (i.e., operational modes). 
Let ${\boldsymbol\mu}_i\in\R_+^{N_i}$ denote a vector of the (approximately) constant power consumption levels of the $i^{th}$ appliance over all of its states. Because each appliance can operate in only one state at a time, we can express the power consumed by appliance $i$ at time $t$ as: 
\begin{equation}\label{eq2}
\begin{aligned}
x_{i,t} = {\bm\mu}_i^T {\bf e}_{i,t}, ~~~\forall \; i \in [L], \; t \in [T]
\end{aligned}
\end{equation}
\noindent where ${\bf e}_{i,t}\in\{0,1\}^{N_i}$ is a binary ``selection'' vector that represents the state of appliance $i$ at time $t$ and whose entries sum to $1${, i.e., $\mathbf{1}^T{\mathbf{e}}_{i,t} = {1}$. }

\subsection{Formulation}
Conditioned on the power consumption profiles $\{{\boldsymbol\mu}_i\}_{i=1}^{L}$ and the connectivity weights $\{w_i^{r}\}_{(i,r=1)}^{(L,R)}$ being known \emph{a priori}, the energy disaggregation problem boils down to choosing a state for each appliance at a time, $t$. Although exploiting the aggregated measurements from multiple lines somewhat reduces the ill-posedness of the problem, from an ``equations versus unknowns'' standpoint, it is always under-determined. Consequently, 
we exploit the fact that appliances change states infrequently over a short time horizon. Hence, 
we propose performing the energy disaggregation task over the entire time horizon while imposing temporal consistency on the evolution of the binary selection vectors. This leads to the following formulation:
\begin{equation} \label{eq:pb_form_binary}
	\begin{aligned}
	 \underset{\{{\bf e}_{i,t}\}_{(i,t=1)}^{(L,T)}}{\min}
	 & ~~ \sum_{r,t=1}^{R,T} 
	 \big({y}^r_t - \sum_{i=1}^{L} {w}_i^{r}\bm{\mu}_i^T\mathbf{e}_{i,t}\big)^2
	 -  \sum_{i,t=1}^{L,T-1} 
	 \lambda_i {\mathbf{e}_{i,t}}^T\mathbf{e}_{i,t+1}\\[0pt]
	\!\!\!{\rm s.t.}~
	 & {\bf e}_{i,t}\in\{0,1\}^{N_i},
	\mathbf{1}^T{\mathbf{e}}_{i,t} = {1}, \forall \; i \in [L],t \in [T]
	\end{aligned} 
\end{equation}
\noindent where the first term represents the least-squares data fit over all phases (lines); and the second term is a smoothness-inducing regularizer that seeks to maximize the similarity between the states of an appliance over consecutive time instants as in \cite{Powerlets,ouricassp}; and  
$\lambda_i \in \mathbb{R}_{+}$ is a regularization parameter (we set it to 1 in the experiments).  
The constraints in \eqref{eq:pb_form_binary} guarantee the selection of only one state for each appliance at a time. 
Evidently, this problem is a discrete quadratic program, which is NP--hard in its general form. As such, our objective is to design an approximation algorithm capable of yielding high-quality, albeit suboptimal solutions
in polynomial time. As a first step, we equivalently reformulate \eqref{eq:pb_form_binary} as a \emph{subset selection} problem. This requires expressing \eqref{eq:pb_form_binary} in set-notation, which is done as follows.

For each appliance $i \in [L]$, we define a ``ground'' set $\setA_i := \{1,...,N_i\}$ that represents the universe of states that appliance $i$ can occupy. Then, let $\mathcal{S}_{i,t}$ be the singleton set that represents the state of appliance $i$ at time $t$. Simple inspection reveals that ${\bf e}_{i,t}$ is the indicator vector of $\setS_{i,t}$, i.e., ${\bf e}_{i,t} = \mathbbm{1}_{\setS_{i,t}}$. 
As an example, if appliance $i$ has $N_i = 4$ states, and it is operating in the third state at $t$, then $\mathbf{e}_{i,t} = [0,0,1,0]^T \leftrightarrow \setS_{i,t} = \{3\}$.
To express the problem concisely, let the set $\setS_t := \bigcupdot_{i=1}^L \setS_{i,t}$ be the \emph{disjoint} union of the sets $\{\setS_{i,t}\}_{i=1}^L$, i.e., $\setS_t$  ``concatenates'' the states of all appliances at $t$ as $\setS_t := [\setS_{1,t},\cdots,\setS_{L,t}]$. 
Analogously, we define the set $\setT:=\bigcupdot_{i=1}^L \setA_{i}$ to be the ``super-universe'' of all states across all appliances. 
Let $N:= \sum_{i=1}^{L} N_i$.
Then, we define:

\begin{equation*}
\bm{\beta}^r := [w_1^r\bm{\mu}_1^T, w_2^r\bm{\mu}_2^T, \dots, w_L^r\bm{\mu}_L^T]^T \in \R^N, \forall r \in [R]
\end{equation*}
\noindent which concatenates the consumption vectors of all appliances connected to line $r$ {\em and} scales them by their respective connectivity weights, $w_i^r$. 
Next, define the matrix $\mathbf{B}^r := {\boldsymbol \beta}^r{{\boldsymbol \beta}^r}^T$ and the vector $\mathbf{b}^r_t = 2y^r_t{\boldsymbol \beta}^r$ for each line $r \in [R]$. Finally, we define the diagonal matrix ${\boldsymbol \Lambda}:=\textrm{diag}(\lambda_1\mathbf{1}_{N_1}, \dots, \lambda_L\mathbf{1}_{N_L})$, where $\mathbf{1}_{N_1}$ is a vector of all ones of size $N_i$. 
Putting everything together and expanding the least-squares terms, \eqref{eq:pb_form_binary} can be equivalently expressed as: 
\begin{equation} \label{eq:pb_time}	
	\begin{aligned}
	 \underset{\{\setS_t \in \setI_t\}_{t=1}^{T}}{\min}
	 \hspace{3mm} \textstyle \textstyle \sum_{r=1}^{R} \sum_{t=1}^{T}
	 \big(\mathbbm{1}_{\setS_t}^T\mathbf{B}^r\mathbbm{1}_{\setS_t} - \mathbbm{1}_{\setS_t}^T{\mathbf{b}^r_t}\big)
	  - \textstyle \sum_{t=1}^{T-1} 
	 \big(\mathbbm{1}_{\setS_t}^T\mathbf{\Lambda}\mathbbm{1}_{\setS_{t+1}}\big)
	\end{aligned} 
\end{equation}
\noindent where the set 
$\setI_t := \{\setS_t \subset \setT: |\setS_t \cap \setA_i| = 1,~\forall \,i \in [L], t \in [T]\}$
guarantees that only one state is chosen per appliance at any time.
To further simplify the problem representation, we define $\setS := \bigcupdot_{t=1}^T\setS_t$ as the set that contains the states of all appliances across all time instants. Note that $\setS \subset \setV := \bigcupdot_{t=1}^T \setT$.
We also define the block diagonal matrix $\mathbf{Q}^r := \mathbf{I}_T \otimes \mathbf{B}^r$, where $\mathbf{I}_T$ is the $T \times T$ identity matrix and $\otimes$ is the Kronecker product.
Next, we define the time smoothness regularization matrix $\mathbf{R} := {\bf D} \otimes \boldsymbol{\Lambda}$, where ${\bf D} \in \mathbb{R}^{T \times T}$ is a symmetric Toeplitz matrix, whose first super- and sub-diagonal elements equal $1/2$, and the remaining entries are $0$.
Finally, let $\mathbf{b}^r := [{\mathbf{b}^r_1}^T, {\mathbf{b}^r_2}^T, \cdots, {\mathbf{b}^r_t}^T]^T$.  Armed with these definitions, we obtain the final subset-selection form of
\eqref{eq:pb_form_binary}:
\begin{equation} \label{eq:eqn3}	
\begin{aligned}
\underset{\setS \in \setI}{\min}
& ~~ \big\{ f(\setS):= 
\textstyle  \sum_{r=1}^{R}
\big(\mathbbm{1}_{\setS}^T\mathbf{Q}^r\mathbbm{1}_{\setS} - \mathbbm{1}_{\setS}^T{\mathbf{b}^r}\big) 
-\mathbbm{1}_{\setS}^T\mathbf{\mathbf{R}}\mathbbm{1}_{\setS}\big\}\\
\end{aligned} 
\end{equation}
\noindent where $\setI := \bigcupdot_{t=1}^T \setI_t$. Although an exact minimization of the quadratic set functions is NP–hard in general, we now demonstrate that the cost function of \eqref{eq:eqn3} exhibits a special property that enables us to devise a simple polynomial-time approximation algorithm{---the proof is deferred to Appendix \ref{Appd:A}.}

\begin{proposition}\label{prob:DSF}
The set function $f(\setS)$ can be equivalently expressed as a DSF: $f(\setS) = g(\setS) - h(\setS)$,
where $g(\setS):=-\mathbbm{1}_{\setS}^T\mathbf{\mathbf{R}}\mathbbm{1}_{\setS}$ and $h(\setS):= \sum_{r=1}^{R} -\mathbbm{1}_{\setS}^T\mathbf{Q}^r\mathbbm{1}_{\setS} + \mathbbm{1}_{\setS}^T{\mathbf{b}^r}$ are submodular functions.
\end{proposition}

\subsection{Algorithm}
To exploit the DSF form in our formulation, we utilize a discrete optimization analogue of the majorization-minimization (MM) procedure proposed in \cite{narasimhan2012submodular,iyer2012algorithms}. The approach is iterative and consists of two main steps:

\noindent 1) {\bf Majorization:} At each iteration $k \in \mathbbm{N}$, we compute a modular {\em upper} bound $u_{\setS^k}^{g}(\setS)$ of $g(\setS)$ about the current solution set $\setS^{k}$ that satisfies the following properties: 
\begin{equation}\label{eq:upper_cond}
 g(\setS) \leq u_{\setS^k}^g(\setS), \forall \setS \subset \setV, ~~\text{and}~ g(\setS^{k})= u_{\setS^k}^g(\setS^k)  
\end{equation}
\noindent Similarly, a modular {\em lower} bound $v_{\setS^k}^h(\setS)$ of $h(\setS)$ is constructed for the current solution set $\setS^{k}$ such that:
\begin{equation}\label{eq:lower_cond}
h(\setS) \geq v_{\setS^k}^h(\setS), \forall \setS \subset \setV, ~~\text{and}~ h(\setS^{k})= v_{\setS^k}^h(\setS^k).    
\end{equation} 
\noindent 2) {\bf Minimization:} Upon replacing $g(\setS)$ by $u_{\setS^k}^g(\setS)$ and $h(\setS)$ by $v_{\setS^k}^h(\setS)$, we obtain  a modular {\em upper} bound of $f(\setS)$, which is tight around the current solution set $\setS = \setS^{k}$. The resulting problem corresponds to minimizing a modular function
\begin{equation} \label{eq:mod_approx}	\begin{aligned}
\underset{\setS \in \setI}{\min}
& ~~ m_k(\setS):= u_{\setS^k}^g(\setS) - v_{\setS^k}^h(\setS)
\end{aligned} 
\end{equation} 
\noindent which admits a simple solution. To see this, note that $m_k(\setS)$ is a modular function by construction, i.e., $m_k(\setS) = {\bf m}_k^T\mathbbm{1}_{\setS}$. 
To compute the optimal solution, we simply inspect the entries of ${\bf m}_k$ corresponding to each subset $\setS_{i,t}$ and pick the index of the smallest entry, $\forall i \in [L], t \in [T]$, which costs only $\mathcal{O}(NT)$ in total. 

\noindent\textbf{Modular Upper Bound:} 
Given a set $\setY\subseteq \setV$, the super-differential set $\partial^g(\setY)$ of a submodular function $g(\setY)$ is defined as \cite{iyer2013fast}: $\partial^g(\setY) = \{ {\bf y}\in \R^n : g(\setX)  \leq  g(\setY) + y(\setX) - y(\setY), \forall \setX \subseteq \setV\}$, where every vector ${\bf y}\in\partial^g(\setY)$ defines a modular function $y(\setX)={\bf y}^T\mathbbm{1}_{\setX}, \forall \setX \subseteq \setV$. A supergradient ${\bf y} \in \partial^h(\setY)$ is used to define a modular upper bound function of the form: $u^g_{\setY}(\setX):= g(\setY) + y(\setX) - y(\setY)$, which, by construction, satisfies the properties \eqref{eq:upper_cond}.
A particular choice of a supergradient ${\bf u}_{\setY}^g \in \partial^g(\setY)$ is given by \cite{iyer2012algorithms}:
\begin{equation}\label{eq:mod_upper}
{\bf u}_{\setY}^g(j) = 
\begin{cases} g(\setY) - g(\setY \backslash \{j\}), &\forall j\in \setY\\
g(\{j\}) - g(\emptyset), &\forall j \notin \setY
\end{cases}
\end{equation}
\noindent With ${\bf u}^g_{\setY}$ obtained, we define the modular function for all subsets $\setS \subseteq \setV$ as $u^g_{\setY}(\setS) = \mathbbm{1}_{\setS}^T  {{\bf u}^g_{\setY}}$,
which we then use as the desired upper bound function in the majorization step.

\noindent\textbf{Modular Lower Bound:} The subdifferential set of a submodular function $h(.)$ for a given set $\mathcal{Y} \subseteq \mathcal{V}$ is defined as \cite[Section 6.2]{fujishige2005submodular}: $\partial_h(\setY) = \{ {\bf y}\in \R^n : h(\setX) \geq h(\setY) + y(\setX) - y(\setY), \forall \setX \subseteq \setV\}$.
Let ${\bf v}_{\setY}^{h} \in \partial_h(\setY)$ denote a subgradient of $h$ at $\setY$. 
We {need} to compute such a subgradient for constructing our desired modular lower bound. To do so, it suffices to compute any element in
the set of extreme points of $\partial h(\setY)$, {which can be 
characterized 
as follows.}
{\begin{proposition}[\cite{fujishige2005submodular}, Theorem 6.11]\label{prop:extreme_points}
For each $\setY \subseteq \setV$, a vector ${\bf v}_{\setY}^{h}$ is an extreme point of $\partial_h(\setY)$ iff there exists a maximal chain $\mathcal C:\emptyset=\setS^{(0)} \subset \setS^{(1)} \subset \cdots\subset \setS^{(n)}=\setV=[n]$ which includes $\setY$ (i.e., $\setY = \setS^{(j)}$ for some $j \in [n]$) such that the modular function $v_{\setY}^{h}$ associated with ${\bf v}_{\setY}^{h}$ satisfies
\begin{equation}\label{eq3.b3}
\begin{split}
v_\setY^h(\setS^{(i)}\backslash\setS^{(i-1)})& = v_\setY^h(\setS^{(i)}) - v_\setY^h(\setS^{(i-1)})\\ &= h(\setS^{(i)}) - h(\setS^{(i - 1)}),\forall \; i \in [n]
\end{split}
\end{equation}	
\end{proposition}
\noindent Using the above description,}
Edmonds \cite{edmonds1970submodular} presented a greedy procedure for computing such extreme points. Given a set $\setY$, let ${\bm \pi}$ be a permutation of the ground set $\setV=[n]$, which maps the elements of $\setY$ to the first $|\setY|$ positions, i.e., $\bm\pi(i) \in \setY, \forall \; i \leq |\setY|$. The remaining $n-|\setY|$ positions of $\bm \pi$ can be assigned randomly. 
Every such permutation vector defines a chain of subsets $\setS_{\bm\pi}^{(0)} \subset \setS_{\bm\pi}^{(1)} \subset ... \subset \setS_{\bm\pi}^{(n)}$ with elements $\setS_{\bm\pi}^{(0)} = \emptyset$, and $\setS_{\bm\pi}^{(i)} = \{\bm\pi(1), \bm\pi(2),\cdots,\bm\pi(i)\}, \forall \; i \in [n]$ ordered by inclusion, i.e., a (maximal) chain. Note that we have $\setS_{\bm\pi}^{|\setY|} = \setY$. Using this chain, we define a vector ${\bf v}_{\setY,\bm{\pi}}^{h} \in \R^n$ with entries:
\begin{equation}\label{eq:mod_lower}
{\bf v}^h_{\setY,\bm{\pi}}(\bm\pi(i)) = 
\begin{cases} h(\setS_{\bm\pi}^{(1)}) &\mbox{if } i=1\\
h(\setS_{\bm\pi}^{(i)}) - h(\setS_{\bm\pi}^{(i-1)}), & \mbox{otherwise}
\end{cases}
\end{equation}
\noindent By construction, ${\bf v}^h_{\setY,\bm{\pi}}$ satisfies the description of an extreme point of $\partial_h(\setY)$ in Theorem 6.11 in  \cite{fujishige2005submodular}. 
With vector ${\bf v}^h_{\setY,\bm{\pi}}$ thus obtained, we define the modular function for all subsets $\setS \subseteq \setV$ as $v^h_{\setY,\bm{\pi}}(\setS) := \mathbbm{1}_{\setS}^T {{\bf v}^h_{\setY,\bm{\pi}}}$.
Further, it has been shown \cite{Grotschel} that for every $\setY \subseteq \setV$, the modular function $v^h_{\setY,\bm{\pi}}(\setS)$ satisfies the following properties: (i) $v^h_{\setY,\bm{\pi}}(\setS) \leq h(\setS), \forall \; \setS \subseteq \setV$, and (ii) $v^h_{\setY,\bm{\pi}}(\setS_{\bm{\pi}}^{(i)}) = h(\setS_{\bm{\pi}}^{(i)}), \forall \; i\in[n]$. While (i) implies the lower bound property, (ii) implies that:
\begin{equation}\label{galg}
v^h_{\setY,\bm{\pi}}(\setS_{\bm{\pi}}^{|\setY|}) = v^h_{\setY,\bm{\pi}}(\setY) = h(\setY).
\end{equation}
\noindent Taken together, the obtained modular function $v^h_{\setY,\bm{\pi}}(\setS)$ is a tight lower bound of the submodular function $h(\setY)$ and satisfies the desired properties in \eqref{eq:lower_cond}. 

The \method algorithm  is summarized in Algorithm \ref{alg:algo1}. 
The procedure exploits the DSF structure of the cost function to perform approximate minimization by successively minimizing a sequence of global upper bounds while respecting the constraints. 
{Regarding the complexity of the subroutines that compute the modular upper and lower bounds at each step, it may appear that these procedures are computationally expensive owing to the fact that evaluating $g(.)$ and $h(.)$ requires instantiating a Kronecker product of potentially large, dense matrices. However, we can show that by means of intelligent manipulation, the sub- and super-gradients can be evaluated using elementary matrix-vector multiplications (see Section \ref{appd:complexity} of the Appendix).}
{Regarding the generated iterates, we have the following claim---see Section \ref{Appd:C} in the Appendix for the proof.
\begin{proposition}\label{prop:cost_value}
Algorithm \ref{alg:algo1} generates a sequence of solution sets $\{\mathcal{S}^k\}_{k\geq0}$ with monotonically non-increasing cost. 
\end{proposition}}
\begin{algorithm}[h]
\raggedright
	\caption{: \method~Algorithm}
	\label{alg:algo1}
	 \textbf{Initialization:} Set $k:=0,\mathcal{S}^{0} \in \setI$ (randomly initialization).\\
	\textbf{Repeat:}
	 1) Generate permutation $\bm{\pi}$ using $\mathcal{S}^{k}$\\
	\textcolor{white}{\bf Repeat:}
	 2) Compute modular upper bound $u^g_{\setS^k}(.)$ of $g(.)$ using \eqref{eq:mod_upper}\\
	 \textcolor{white}{\bf Repeat:}
	3) Compute modular lower bound $v^h_{\setS^k,\bm{\pi}}(.)$ of $h(.)$ using \eqref{eq:mod_lower}\\
		\vspace{-2pt}
\textcolor{white}{\bf Repeat:}
	4)  Compute $\mathcal{S}^{k+1} \in \arg\underset{\mathcal{S}_t \in \mathcal{I}}{\min}~m_k(\mathcal{S})$ in \eqref{eq:mod_approx} via linear scan \\
		\vspace{-4pt}
\textcolor{white}{\bf Repeat:}
	5) Set $k := k+1$.\\
	\vspace{-2pt}
	\textbf{Until} stopping criterion is met 
\end{algorithm}

\noindent {\bf Learning Connectivity and State Variables:} 
The power profiles $\{\bm{\mu}_i\}_{i=1}^L$ are learned by performing the Lloyd-Max quantization on the power consumption sequence $\{x_i(t)\}_{t=1}^{T}$ in the training data, and setting $\bm{\mu}_i$ to be the centroid values of the quantization intervals. The number of quantization intervals equals the number of states, $N_i$, which is fixed beforehand. 
Thereafter, the connectivity weights, $w_i^r$, are obtained by solving the following convex optimization problem:
\begin{equation} \label{eq:connect_var}	
	\begin{aligned}
	  \underset{\{{w}^r_{i}\}_{(i,r=1)}^{(L,R)}}{\min}
	 & ~~ \textstyle \sum_{r=1}^{R} \textstyle \sum_{t=1}^{T} 
	 \big({y}^r(t) - \textstyle \sum_{i=1}^{L} {w}_i^{r}x_{i,t}\big)^2\\
		\!\!\!{\rm s.t.}~
	 & ~~ 0 \leq w_i^r \leq 1, ~~~\textstyle \sum_{r=1}^R w_i^r = 1, \forall \;i \in [L], r \in [R]
\end{aligned} 
\end{equation}
{\noindent where, $x_{i,t}=\bm{\mu}_i^T{\bf e}_i(t)$ is the appliance-level power as in \eqref{eq2}.}

\section{Experiments}

\subsection{Datasets}
We evaluate \method~using two publicly available datasets: {\bf REDD} and {\bf ECO}. 
Each dataset represents one of the two power distribution systems described earlier as they were collected in homes on different continents. 
{\bf REDD} \cite{kolter2011redd} contains data from 6 homes in the United States (House 5 is omitted because  it does not have enough data).
The whole-home measurements consist of the power readings at two lines; hence, the structure of the distribution system is {\em split-phase}.  
{\bf ECO} \cite{beckel2014nilm} contains data from 6 Swiss households (we omit House 3 because it does not have enough data after synchronizing the time series). The distribution system here is {\em three-phase}, and the aggregated power consumption of each household is available for each phase feeding the premises. 
For both datasets, we collect all the time-stamped readings {that have both the} aggregated \emph{and} appliance-level measurements to ensure synchronized readings, then we down-sample to 1 reading/minute.

\subsection{Baselines and metric}
We compare \method~to four quite different baselines to ensure broad evaluation.
The baseline methods (explained in Section \ref{sec:intro}) are: (i)~{\bf DSC} ({d}iscriminative {s}parse {c}oding) \cite{kolter2010energy}, (ii)~{\bf NMF} \cite{rahimpour2017non}, (iii)~{\bf seq2p} \cite{zhang2018sequence}, and (iv)~{\bf BSMA} ({b}lock {s}uccessive {m}odular {a}pproximation) \cite{ouricassp}.  
We measure the percentage of energy deviated ($PED$) from the true consumption of appliance $i$ in a house $h$ at a time $t$ using:
\begin{equation}\label{eq:RMSE}
	\begin{aligned}
	PED_i(t,h) &:= \frac
	{|{x_{i}(t,h) - \hat{x}_{i}(t,h)}|}{{y{(t,h)}}},
	\end{aligned}
\end{equation}
\noindent where $x_{i}(t,h)$ and $\hat{x}_{i}(t,h)$ are the true and inferred power consumption for appliance $i$ at time $t$ in house $h$, and $y(t,h)$ is the aggregated power at $t$ in $h$. 
Then, we present the average of $PED$ ($\boldsymbol{APED}$) over the total time ticks in all the houses:
\begin{equation}\label{eq:RMSE2}
	\begin{aligned}
	APED_i(t,h) &:= \frac
	{\textstyle \sum_{t=1}^{T} \textstyle \sum_{h=1}^{H} PED_i(t,h)}{\textstyle \sum_{h=1}^{H} T_h}
	\end{aligned}
\end{equation}
\noindent where $T_h$ is the length of the time series of house $h$. The essence of this metric is adopted from \cite{batra2018transferring}. The percentage of energy correctly allocated \cite{kolter2011redd} is a complementary measure that can be represented as ($1-PED$). 
We split the data for each home into two halves---one for training and the other for testing. Note that our approach and the BSMA baseline are optimization-based and do not require training a model; thus, the training data are used only to choose  
the state vectors, $\boldsymbol{\mu}_i$, the number of states, $N_i$, and the connectivity weights, $w_i^r$. 
\begin{table*}[t]
	\caption{$APED\%$ of appliances in REDD and ECO (lower is better). Underline bold means best, bold is second best.}
	\centering
	\label{table1}
	\resizebox{0.9\textwidth}{!}{
	\begin{tabular}{c| c c c c c |c|  c c c c c}
		\hline
		  \multicolumn{6}{c|}{\bf{REDD}} & \multicolumn{6}{|c}{\bf{ECO}}\\
		  \hline	
		Appliance &  DSC&  NMF  &  seq2p  &  BSMA  &  \method  &  Appliance  &  DSC  &  NMF  &  seq2p  &  BSMA &  \method\\

		\hline		
	    Fridge &  33.72 &  32.32    & \underline{\bf 16.71} &  20.96 & {\bf 20.17}  &
		 Fridge  &  21.53 & {\bf 12.58}     &     \underline{\bf 11.75}         &     14.00    &  13.85    \\
		 Dishwasher  &  3.97 & 5.47    & 5.17 &  {\bf 2.98}  &  \underline{{\bf 2.22}} &
		    Dishwasher    &  5.63   &      16.85      &    18.91        &     {\bf 2.72}     &    \underline{{\bf 2.56}}     \\
    	Microwave  &  3.32  & 3.21    & 9.76  &   {\bf 3.12}  & \underline{{\bf 2.84}} &
		 Microwave      &   12.78    &    15.21       &     {\bf 4.74}       &    7.03     &   \underline{{\bf 3.57}}   \\
		Washer/dryer  &  10.23 & 13.93    & {\bf 2.66}  &   {\bf 2.66} & \underline{{\bf 1.79}} &
		\hspace{-2mm}   Washer/dryer  \hspace{-2mm}  &    30.53  &  {\bf  2.58}       &    3.21        &   2.84        & \underline{{\bf 0.90}}   \\
    	Stove  &  4.94 &  4.46    & \underline{{\bf 1.62}} &  4.02 & {\bf 1.75}  &
    	 Stove      &   2.11   &      1.65     &        7.41      &    {\bf   0.63 }   &    \underline{{\bf 0.53}}   \\
		AC  &  1.80 & \underline{{\bf 1.57}}    & 1.74 & 1.86 &  {\bf 1.64} &
	    Freezer     &  26.74   &     22.31     &     {\bf 17.00}         &     \underline{{\bf 18.56}}    &   25.06    \\ 	
	    	Bathroom GFI  &   4.61 & 5.35    & 3.00 &  {\bf 1.01} & \underline{{\bf 0.71}} &
		 Work station &   31.05   &     11.62       &    \underline{{\bf 3.00} }       &   6.77      &  {\bf 6.68}     \\ 
		\hspace{-2mm} Outlet unknown \hspace{-2mm}  &  6.72  & 8.23    & \underline{{\bf 2.63}}  &  9.85 & {\bf 4.94}  &
		 TV $\&$ stereo    &  17.45   &     12.33      &    \underline{\bf 5.91} &    16.28     &   {\bf 10.96}   \\
	  Kitchen outlet &  13.76 & 15.03     & {\bf 5.43} &  6.18 &      \underline{{\bf 5.33}}  &
		 Tablets      &   19.50   &     12.25      &     0.47    &    \underline{{\bf 0.58} }      &   \underline{{\bf 0.58}}    \\
        Lighting  &  19.21 & 17.79    & \underline{{\bf 5.53}} &   12.31  & {\bf 9.12} &
		  --     &  --   &      --    &      --      &    --     &   --   \\
		\hline
		{\bf Average} & 10.23 & 10.73  & {\bf 5.42} &  6.49 & \underline{\bf 5.05} &
		{\bf Average}  &    18.59  &      11.93     &      8.04        &      {\bf 7.71}     &   \underline{{\bf 7.19}}   \\
		\hline
	\end{tabular}
	}
	\vspace*{-2mm}
\end{table*}

\subsection{Results}
Table \ref{table1} 
shows
the prediction error for each appliance in the REDD and ECO data---we show appliances that appears three times or more. The homes in ECO do not have consistent types of appliances; thus, we also include the typical appliances (e.g., microwave, stove) in Table \ref{table1} in addition to the common ones among households. 
With the REDD data, \method has four appliances with the $APED$ less than $2\%$, whereas all the baselines have only two appliances less than $2\%$.  
Compared to the baselines, \method significantly improves the prediction of appliances. \method reduces the average of the $APED$ among all appliances with DSC, NMF, seq2p, and BSMA by $50.6\%$, $52.9\%$, $6.88\%$, and $22.2\%$, respectively, on the REDD data.
\method also improves the mean of the $APED$ among DSC, NMF, seq2p, and BSMA using the ECO data by $61.34\%$, $39.75$, $10.63\%$, and $6.79\%$, respectively.    
Moreover, \method has the best (or comparable in a few cases) performance for appliances with heavier load (e.g., washer/dryer, AC, fridge, and stove) and appliances with flexible usage time, e.g., dishwasher. 
Note that in a recent survey study \cite{batra2019towards}, seq2p has been shown to be the strongest baselines with heavy load appliances.
For instance, the $APED$ of \method with washer/dryer in the REDD data is only $17.45\%$, $12.83\%$, $67.16\%$, and $67.24\%$ of the $APED$ of DSC, NMF, seq2p, and BSMA, respectively. With the ECO data, the washer/dryer error percentage of our method is only $2.96\%$, $34.98\%$, $28.13\%$, and $31.85\%$ of the $APED$ of the DSC, NMF, seq2p, and BSMA, respectively. 

\section{Conclusions}
In this paper, we presented a supervised framework for energy disaggergation that exploits the structure of power distribution systems by using multiple aggregated measurements to improve the disaggregation accuracy. The proposed approach formulates the problem as minimizing the difference between two submodular functions, subject to combinatorial constraints. Leveraging this form, we devised an iterative approximation algorithm that minimizes a sequence of global modular upper bounds on the cost function.  The algorithm provably exhibits a non-increasing cost and features computionally lightweight updates. The effectiveness of \method~was shown against four state-of-the-art baselines on two datasets with different power connectivity structures.

\appendices
\def \va{{\mathbf a}}
\def \vb{{\mathbf b}}
\def \vc{{\mathbf c}}
\def \vd{{\mathbf d}}
\def \ve{{\mathbf e}}
\def \vh{{\mathbf h}}
\def \vg{{\mathbf g}}
\def \vm{{\mathbf m}}
\def \vn{{\mathbf n}}
\def \vp{{\mathbf p}}
\def \vq{{\mathbf q}}
\def \vr{{\mathbf r}}
\def \vs{{\mathbf s}}
\def \vu{{\mathbf u}}
\def \vv{{\mathbf v}}
\def \vw{{\mathbf w}}
\def \vx{{\mathbf x}}
\def \vy{{\mathbf y}}
\def \vz{{\mathbf z}}
\def \vA{{\mathbf A}}
\def \vB{{\mathbf B}}
\def \vC{{\mathbf C}}
\def \vD{{\mathbf D}}
\def \vE{{\mathbf E}}
\def \vF{{\mathbf F}}
\def \vG{{\mathbf G}}
\def \vH{{\mathbf H}}
\def \vI{{\mathbf I}}
\def \vJ{{\mathbf J}}
\def \vK{{\mathbf K}}
\def \vP{{\mathbf P}}
\def \vQ{{\mathbf Q}}
\def \vS{{\mathbf S}}
\def \vM{{\mathbf M}}
\def \vR{{\mathbf R}}
\def \vU{{\mathbf U}}
\def \vV{{\mathbf V}}
\def \vW{{\mathbf W}}
\def \vX{{\mathbf X}}
\def \vZ{{\mathbf Z}}

\section{Proof of Proposition \ref{prob:DSF}}\label{Appd:A}
\begin{proof}
Define $g(\setS):= -\mathbbm{1}_{\setS}^T\mathbf{\mathbf{R}}\mathbbm{1}_{\setS}$ and $h(\setS):= \sum_{r=1}^{R} -\mathbbm{1}_{\setS}^T\mathbf{Q}^r\mathbbm{1}_{\setS} + \mathbbm{1}_{\setS}^T{\mathbf{b}^r}$. Evidently, both $g(.)$ and $h(.)$ are quadratic set functions. A necessary and sufficient condition for such a set function to be submodular is the following:
\begin{lemma}[\cite{bach2013learning}, Proposition 6.3]\label{lemma:1}
A quadratic set function $f: \setX \rightarrow \mathbbm{1}_{\setX}^T\mathbf{\mathbf{A}}\mathbbm{1}_{\setX}$ is submodular if and only if all off-diagonal elements of $\mathbf{A}$ are non-positive.
\end{lemma}

We now establish that this condition is satisfied by both set functions $g(.)$ and $h(.)$. First, consider $g(\setS)$, by construction, the off-diagonal elements of $-\mathbf{R}$ are non-positive because $\lambda_i \geq 0 \Rightarrow \mathbf{\Lambda} \geq 0 \rightarrow -\mathbf{R} \leq 0$. Thus, $g(\setS)$ is submodular via Lemma~\ref{lemma:1}. Similarly, we can show that all off-diagonal elements in $-\mathbf{Q}^r$ are non-positive because $w_i^r \geq 0$ and $\bm{\mu}_i\geq0 \Rightarrow \bm{\beta}^r \geq 0 \Rightarrow \mathbf{B}^r \geq 0 \Rightarrow -\mathbf{Q}^r \leq 0$, making $-\mathbbm{1}_{\setS}^T\mathbf{Q}^r\mathbbm{1}_{\setS}$ submodular. Meanwhile, the second term in $h(\setS)$ $\mathbbm{1}_{\setS}^T{\mathbf{b}^r}$ is modular. Since the class of submodular functions is closed under non-negative linear combinations, $h(\setS)$ is also submodular.  
\end{proof}

\section{Complexity Analysis}\label{appd:complexity}

In this section, we provide a detailed analysis of the computational cost incurred in computing a subgradient of $h(\setS)$ \eqref{eq:mod_lower} and supergradient of $g(\setS)$ \eqref{eq:mod_upper}. We leverage the Kronecker product form of both quadratic set functions to demonstrate that these key subroutines (required at each step of Algorithm 1) can be carried out using only simple matrix-vector multiplications.

\subsection{Subgradient computation:}
The set function $h(\setS)$ has the form:
\begin{equation}
    h(\setS) =   -\mathbbm{1}_{\setS}^T(\sum_{r=1}^{R}\mathbf{Q}^r)\mathbbm{1}_{\setS} + (\sum_{r=1}^{R}{\mathbf{b}^r})^T\mathbbm{1}_{\setS}
\end{equation}
For a given set $\setY \subseteq \setV$ and an appropriate permutation of the ground set $\bm{\pi} \in [NT]$, a subgradient of $h(.)$ at $\setY$ has the form \eqref{eq:mod_lower}:
\begin{equation}
{\bf v}^h_{\setY,\bm{\pi}}(\bm\pi(i)) = 
\begin{cases} 
h(\setS_{\bm\pi}^{(1)}) &\mbox{if } i=1\\
h(\setS_{\bm\pi}^{(i)}) - h(\setS_{\bm\pi}^{(i-1)}), & \mbox{otherwise}
\end{cases}
\end{equation}
For $i=1$, we have: 
\begin{equation}\label{eq:i1}
h(\setS_{\bm\pi}^{(1)}) = -\mathbf{e}^T_{\setS_{\bm\pi}^{(1)}}
(\sum_{r=1}^{R} \vI_T \otimes \vB^{r})\mathbf{e}_{\setS_{\bm\pi}^{(1)}} + (\sum_{r=1}^{R}{\mathbf{b}^r})^T\mathbf{e}_{\setS_{\bm\pi}^{(1)}}
\end{equation}
where $\mathbf{e}_{\setS_{\bm\pi}^{(1)}}$ is a binary indicator vector of the singleton set $\setS_{\bm\pi}^{(1)}$. Let $\vZ \in \{0,1\}^{N \times T}$ be a matrix such that $\mathbf{e}_{\setS_{\bm\pi}^{(1)}} = \text{vec}(\vZ)$. Then,  using the linearization property of the $\text{vec}(.)$ operator and the cyclic property of the trace operator, we obtain:
\begin{equation}\label{eq:vec}
\begin{aligned}
h(\setS_{\bm\pi}^{(1)}) &= -\sum_{r=1}^{R}(\text{vec}(\vZ)^T\textrm{vec}(\vB^{r}\vZ\vI_T)) + (\sum_{r=1}^{R}{\mathbf{b}^r})^T \text{vec}(\vZ)\\
&= -\sum_{r=1}^{R} \text{trace}(\vZ^T\vB^r\vZ) + (\sum_{r=1}^{R}{\mathbf{b}^r})^T \text{vec}(\vZ)
= - \text{trace}(\vZ^T\Bar{\vB}^r\vZ) +  \text{vec}(\Bar{\vC})^T\text{vec}(\vZ)  
\end{aligned}
\end{equation}
where we have defined $\Bar{\vB}^r := \sum_{r=1}^{R}{\mathbf{B}^r}$ and $\Bar{\vC} \in \mathbb{R}^{N \times T}$ such that $\sum_{r=1}^{R}{\mathbf{b}^r} = \text{vec}(\Bar{\vC})$. Since the vectorization of $\vZ$ produces a binary indicator vector, it follows that $\vZ$ admits a rank-$1$ decomposition of the form $\vZ = \ve_n\ve_t^T$ for some canonical basis vectors $\ve_n$ and $\ve_t$ of $\mathbb{R}^N$ and $\mathbb{R}^T$ respectively (here $n \in [N]$ and $t \in [T]$ respectively). This allows us to further simplify the terms of \eqref{eq:vec} as follows:
\begin{equation}
\begin{aligned}
\text{trace}(\vZ^T\Bar{\vB}^r\vZ) &= \text{trace}(\ve_t\ve_n^T\Bar{\vB}^r\ve_n\ve_t^T) = \ve_n^T\Bar{\vB}^r\ve_n = \sum_{r=1}^R \ve_n^T(\bm{\beta}^r)(\bm{\beta}^r)^T\ve_n 
= \sum_{r=1}^R(\ve_n^T\bm{\beta}^r)^2 
= \sum_{r=1}^R \bm{\beta}^r(n)^2,\\
\text{vec}(\Bar{\vC})^T\text{vec}(\vZ) &= \text{trace}(\Bar{\vC}^T\vZ) 
= \text{trace}(\Bar{\vC}^T\ve_n\ve_t^T)
= \ve_n^T\Bar{\vC}\ve_t =  \Bar{\vC}(n,t)
\end{aligned}
\end{equation}
Hence, we have: 
\begin{equation}\label{eq:final1}
  h(\setS_{\bm\pi}^{(1)}) = - \sum_{r=1}^R \bm{\beta}^r(n)^2 + \Bar{\vC}(n,t).
\end{equation}
For the general case of $i \geq 2$, we can express each term ${\bf v}^h_{\setY,\bm{\pi}}(\bm\pi(i))$ as: 
\begin{equation}\label{eq:general}
    {\bf v}^h_{\setY,\bm{\pi}}(\bm\pi(i)) = 
    \underbrace{\biggl[\mathbf{e}^T_{\setS_{\bm\pi}^{(i-1)}}
(\sum_{r=1}^{R} \vI_T \otimes \vB^{r})\mathbf{e}_{\setS_{\bm\pi}^{(i-1)}}
    -\mathbf{e}^T_{\setS_{\bm\pi}^{(i)}}
(\sum_{r=1}^{R} \vI_T \otimes \vB^{r})\mathbf{e}_{\setS_{\bm\pi}^{(i)}}\biggr]}_{(\text{A})}
+\underbrace{\biggl[(\sum_{r=1}^{R}{\mathbf{b}^r})^T(\mathbf{e}_{\setS_{\bm\pi}^{(i)}} - \mathbf{e}_{\setS_{\bm\pi}^{(i-1)}})}_{(\text{B})}
\biggr]
\end{equation}
where $\mathbf{e}_{\setS_{\bm\pi}^{(i)}}$ denotes the binary indicator vector of the set ${\setS_{\bm\pi}^{(i)}}$. 
By construction, ${{\setS_{\bm\pi}^{(i-1)}} \subset \setS_{\bm\pi}^{(i)}}$ and $|{\setS_{\bm\pi}^{(i)}} \backslash {\setS_{\bm\pi}^{(i-1)}}| = 1$. Hence, it follows that the vector $\mathbf{e}_{\setS_{\bm\pi}^{(i)}}$ can be expressed as 
$\mathbf{e}_{\setS_{\bm\pi}^{(i)}} = \mathbf{e}_{\setS_{\bm\pi}^{(i-1)}} +  \mathbf{e}_{\bm\pi(i)}$. This observation allows us to simplify the constituent terms of \eqref{eq:general} as follows:
\begin{equation}\label{eq:termA}
(\text{A}) = -\mathbf{e}_{\bm\pi(i)}^T(\sum_{r=1}^{R} \vI_T \otimes \vB^{r}) \mathbf{e}_{\bm\pi(i)} -2\mathbf{e}_{\bm\pi(i)}^T(\sum_{r=1}^{R} \vI_T \otimes \vB^{r}) 
{\mathbf{e}}_{\setS_{\bm\pi}^{(i-1)}}
\end{equation}
Note that the first term in the above equation is similar to the quadratic term in \eqref{eq:i1}, and by employing a similar reasoning as illustrated before, we obtain:
\begin{equation}\label{eq:termA1}
\mathbf{e}_{\bm\pi(i)}^T(\sum_{r=1}^{R} \vI_T \otimes \vB^{r}) \mathbf{e}_{\bm\pi(i)}
=  \text{trace}(\vZ_{\bm\pi(i)}^T\Bar{\vB}^r\vZ_{\bm\pi(i)}) =  \sum_{r=1}^R \bm{\beta}^r(n_{\bm\pi(i)})^2
\end{equation}
where, by a similar analogy as before, we have defined $\vZ_{\bm\pi(i)} := {\ve_n}_{\bm\pi(i)}{\ve_t}_{\bm\pi(i)}^T$ such that $\mathbf{e}_{\bm\pi(i)} = \text{vec}(\vZ_{\bm\pi(i)})$. Meanwhile, utilizing the linearization property of the $\text{vec}(.)$ operator once again enables us to express the second term in \eqref{eq:termA} as:
\begin{equation}\label{eq:termA2}
\begin{aligned}
 \mathbf{e}_{\bm\pi(i)}^T(\sum_{r=1}^{R} \vI_T \otimes \vB^{r}) \mathbf{e}_{\setS_{\bm\pi}^{(i-1)}} 
 &= \text{vec}(\vZ_{\bm\pi(i)})^T\text{vec}(\Bar{\vB}^r\vZ_{\setS_{\bm\pi}^{(i-1)}}) 
 = \text{trace}(\vZ_{\bm\pi(i)}^T\Bar{\vB}^r\vZ_{\setS_{\bm\pi}^{(i-1)}})
 = \sum_{r=1}^R\text{trace}({\ve_t}_{\bm\pi(i)}{\ve_n}_{\bm\pi(i)}^T  (\bm{\beta}^r)(\bm{\beta}^r)^T\vZ_{\setS_{\bm\pi}^{(i-1)}})\\
 &= \sum_{r=1}^R \bm{\beta}^r(n_{\bm\pi(i)})(\bm{\beta}^r)^T\vZ_{\setS_{\bm\pi}^{(i-1)}}{\ve_t}_{\bm\pi(i)}
 = \sum_{r=1}^R \bm{\beta}^r(n_{\bm\pi(i)})(\bm{\beta}^r)^T\vZ_{\setS_{\bm\pi}^{(i-1)}}(:,t_{\bm\pi(i)})
\end{aligned}
\end{equation}
Finally, the second term of \eqref{eq:general} can be expressed as: 
\begin{equation}\label{eq:termB}
    (\text{B}) = (\sum_{r=1}^{R}{\mathbf{b}^r})^T\mathbf{e}_{\bm\pi(i)}
    = \text{vec}(\Bar{\vC})^T\text{vec}(\vZ_{\bm\pi(i)}) = \text{trace}(\Bar{\vC}^T\vZ_{\bm\pi(i)}) 
    = \text{trace}(\Bar{\vC}^T{\ve_n}_{\bm\pi(i)}{\ve_t}_{\bm\pi(i)}^T) = \Bar{\vC}(n_{\bm\pi(i)},t_{\bm\pi(i)})
\end{equation}
Combining \eqref{eq:termA}--\eqref{eq:termB}, we obtain:
\begin{equation}\label{eq:final2}
{\bf v}^h_{\setY,\bm{\pi}}(\bm\pi(i)) = -\sum_{r=1}^R \bm{\beta}^r(n_{\bm\pi(i)})^2
-2\sum_{r=1}^R \bm{\beta}^r(n_{\bm\pi(i)})(\bm{\beta}^r)^T\vZ_{\setS_{\bm\pi}^{(i-1)}}(:,t_{\bm\pi(i)}) + \Bar{\vC}(n_{\bm\pi(i)},t_{\bm\pi(i)}), \forall \;i \geq 2
\end{equation}
Hence, from \eqref{eq:final1} and \eqref{eq:final2}, the final expression of the subgradient of the set function $h(.)$ at a given set $\setY \in \setV$ and a permutation vector $\bm\pi \in [NT]$ is:
\begin{equation}
{\bf v}^h_{\setY,\bm{\pi}}(\bm\pi(i)) = 
\begin{cases} 
- \sum_{r=1}^R \bm{\beta}^r(n)^2 + \Bar{\vC}(n,t) &\mbox{if } i=1\\
-\sum_{r=1}^R \bm{\beta}^r(n_{\bm\pi(i)})^2
-2\sum_{r=1}^R \bm{\beta}^r(n_{\bm\pi(i)})(\bm{\beta}^r)^T\vZ_{\setS_{\bm\pi}^{(i-1)}}(:,t_{\bm\pi(i)}) + \Bar{\vC}(n_{\bm\pi(i)},t_{\bm\pi(i)}), & \mbox{otherwise}
\end{cases}
\end{equation}
The expressions reveal that no Kronecker products are needed; instead only simple summations and inner products suffice to compute a subgradient. 

\subsection{Supergradient computation:}
The set function $g(\setS)$ has the form: 
\begin{equation}
g(\setS) = -\mathbbm{1}_{\setS}^T\vR\mathbbm{1}_{\setS} = -\mathbbm{1}_{\setS}^T(\vD \otimes \bm{\Lambda})\mathbbm{1}_{\setS}
\end{equation}
For a given set $\setY \subseteq \setV$, a supergradient of $g(.)$ at $\setY$ has the form \eqref{eq:mod_upper}:
\begin{equation}
{\bf u}_{\setY}^g(j) = 
\begin{cases} g(\setY) - g(\setY 
\backslash\{j\}), &\forall j\in \setY\\
g(\{j\}) - g(\emptyset), &\forall j \notin \setY
\end{cases}
\end{equation}
Note that $g(\emptyset) = 0$, and $g(\{j\}) = -\mathbbm{1}_{\{j\}}^T(\vD \otimes \bm{\Lambda})\mathbbm{1}_{\{j\}}
$. Let $\vZ_j \in \{0,1\}^{N \times T}$ be a matrix such that $\mathbbm{1}_{\{j\}} = \text{vec}(\vZ_j) $. Then, we have:
\begin{equation}
g(\{j\}) = -\text{vec}(\vZ_j)^T(\vD \otimes \bm{\Lambda})\text{vec}(\vZ_j)
= -\text{vec}(\vZ_j)^T\text{vec}(\bm{\Lambda}\vZ_j\vD)
= -\text{trace}(\vZ_j^T\bm{\Lambda}\vZ_j\vD)
\end{equation}
Using the fact that $\vZ_j$ can be decomposed as $\vZ_j = \ve_n\ve_t^T$, for some canonical basis vectors $\ve_n \in \mathbb{R}^N$ and $\ve_t \in \mathbb{R}^T$, 
\begin{equation}
g(\{j\}) = -\text{trace}(\ve_t\ve_n^T\bm{\Lambda}\ve_n\ve_t^T\vD) = -(\ve_n^T\bm{\Lambda}\ve_n)(\ve_t^T\vD\ve_t) = 0,
\end{equation}
since all the diagonal entries of $\vD$ are zeros. Hence, we obtain:
\begin{equation}
 {\bf u}_{\setY}^g(j) = g(\{j\}) - g(\emptyset) = 0, \forall j \notin \setY.
\end{equation}
On the other hand:
$$g(\setY) = -\mathbbm{1}_{\setY}^T(\vD \otimes \bm{\Lambda})\mathbbm{1}_{\setY} 
= -\text{trace}(\vZ_{\setY}^T\bm{\Lambda}\vZ_{\setY}\vD), $$
$$ g(\setY \backslash \{j\}) = -\mathbbm{1}_{\setY \backslash \{j\}}^T(\vD \otimes \bm{\Lambda})\mathbbm{1}_{\setY \backslash \{j\}}
=-\text{trace}((\vZ_{\setY}-\vZ_{j})^T\bm{\Lambda}(\vZ_{\setY}-\vZ_j)\vD)
$$
where $\vZ_{\setY} \in \{0,1\}^{N \times T}$ is a matrix such that $\mathbbm{1}_{\setY} = \text{vec}(\vZ_{\setY})$. Then:
\begin{equation}
\begin{aligned}
g(\setY) - g(\setY\backslash\{j\}) &=  \text{trace}((\vZ_{\setY}-\vZ_{j})^T\bm{\Lambda}(\vZ_{\setY}-\vZ_j)\vD)
-\text{trace}(\vZ_{\setY}^T\bm{\Lambda}\vZ_{\setY}\vD)\\
&= -2\text{trace}(\ve_t^T\ve_n^T\bm{\Lambda}\vZ_{\setY}\vD) + \text{trace}(\ve_t\ve_n^T\bm{\Lambda}\ve_n\ve_t^T\vD)
\end{aligned}
\end{equation}
We have already shown that the last term is zero. The first term can be further simplified as: 
\begin{equation}
\text{trace}(\ve_t^T\ve_n^T\bm{\Lambda}\vZ_{\setY}\vD)
= \ve_n^T\bm{\Lambda}\vZ_{\setY}\vD\ve_t = \bm{\Lambda}(n,:)\vZ_{\setY}{\bf D}(:,t)
\end{equation}
We are now ready to write the final form of the supergradient ${\bf u}_{\setY}^g$:
\begin{equation}
{\bf u}_{\setY}^g(j) = 
\begin{cases} -2\bm{\Lambda}(n,:)\vZ_{\setY}{\bf D}(:,t), &\forall j\in \setY\\
0, &\forall j \notin \setY
\end{cases}
\end{equation}
Again, no Kronecker products are required to be computed. From the final form above, we observe that if the supergradient is computed about any feasible set $\setY \subseteq \setI$, then only $LT$ entries of ${\bf u}_{\setY}^g$ are non-zero. Furthermore, these non-zero values can be obtained via elementary matrix-vector multiplications.

\section{Proof of Proposition \ref{prop:cost_value}}\label{Appd:C}

\begin{proof}
Follows from the following chain of inequalities 
\begin{subequations}
 \begin{alignat}{7}
 f(\mathcal{S}^{k+1}) &= g(\mathcal{S}^{k+1}) - h(\mathcal{S}^{k+1})\\
 & \leq u^g_{\setS^{k}}(\setS^{k+1}) - v_{\mathcal{S}^{k},\bm{\pi}}^h(\mathcal{S}^{k+1})\label{37:2}\\
  & \leq u^g_{\setS^{k}}(\setS^{k}) - v_{\mathcal{S}^{k},\bm{\pi}}^h(\mathcal{S}^{k})\label{37:3}\\
 &= g(\mathcal{S}^{k}) - h(\mathcal{S}^{k})\label{37:4}
 \end{alignat}
\end{subequations}
 where the first inequality in \eqref{37:2} follows from the fact that $g(\mathcal{S}^{k+1}) \leq u_{\mathcal{S}^{k}}^g(\mathcal{S}^{k+1})$ and $h(\mathcal{S}^{k+1}) \geq v_{\mathcal{S}^{k},\bm{\pi}}^h(\mathcal{S}^{k+1})$; and the second inequality \eqref{37:3} stems from the optimality of solving for $\mathcal{S}^{k+1}$; and the last equality \eqref{37:4} is due to the tightness of the modular approximation at $\mathcal{S}=\mathcal{S}^k$.
\end{proof}

\bibliography{ref_new}

\begin{thebibliography}{10}
\providecommand{\url}[1]{#1}
\csname url@samestyle\endcsname
\providecommand{\newblock}{\relax}
\providecommand{\bibinfo}[2]{#2}
\providecommand{\BIBentrySTDinterwordspacing}{\spaceskip=0pt\relax}
\providecommand{\BIBentryALTinterwordstretchfactor}{4}
\providecommand{\BIBentryALTinterwordspacing}{\spaceskip=\fontdimen2\font plus
\BIBentryALTinterwordstretchfactor\fontdimen3\font minus
  \fontdimen4\font\relax}
\providecommand{\BIBforeignlanguage}[2]{{%
\expandafter\ifx\csname l@#1\endcsname\relax
\typeout{** WARNING: IEEEtran.bst: No hyphenation pattern has been}%
\typeout{** loaded for the language `#1'. Using the pattern for}%
\typeout{** the default language instead.}%
\else
\language=\csname l@#1\endcsname
\fi
#2}}
\providecommand{\BIBdecl}{\relax}
\BIBdecl

\bibitem{hart1992nonintrusive}
G.~W. Hart, ``Nonintrusive appliance load monitoring,'' \emph{Proceedings of
  the IEEE}, vol.~80, no.~12, pp. 1870--1891, 1992.

\bibitem{shin2019data}
C.~Shin, S.~Rho, H.~Lee, and W.~Rhee, ``Data requirements for applying machine
  learning to energy disaggregation,'' \emph{Energies}, vol.~12, no.~9, pp.
  1--19, 2019.

\bibitem{faustine2017survey}
A.~Faustine, N.~H. Mvungi, S.~Kaijage, and K.~Michael, ``A survey on
  non-intrusive load monitoring methodies and techniques for energy
  disaggregation problem,'' \emph{arXiv preprint arXiv:1703.00785}, 2017.

\bibitem{parson2015dataport}
O.~Parson, G.~Fisher, A.~Hersey, N.~Batra, J.~Kelly, A.~Singh, W.~Knottenbelt,
  and A.~Rogers, ``Dataport and nilmtk: A building data set designed for
  non-intrusive load monitoring,'' in \emph{Proceedings of the IEEE Global
  Conference on Signal and Information Processing (GlobalSIP)}, Orlando,
  Florida, United States, Dec. 2015, pp. 210--214.

\bibitem{kolter2010energy}
J.~Z. Kolter, S.~Batra, and A.~Y. Ng, ``Energy disaggregation via
  discriminative sparse coding,'' in \emph{Proceedings of the Advances in
  Neural Information Processing Systems (NIPS)}, Vancouver, British Columbia,
  Canada, Dec. 2010, pp. 1153--1161.

\bibitem{Powerlets}
E.~Elhamifar and S.~Sastry, ``Energy disaggregation via learning
  ‘powerlets’ and sparse coding,'' in \emph{Proceedings of the 29th AAAI
  Conference on Artificial Intelligence}, Austin, Texas, United States, Jan.
  2015, pp. 629--635.

\bibitem{pandey2019structured}
S.~Pandey and G.~Karypis, ``Structured dictionary learning for energy
  disaggregation,'' in \emph{Proceedings of the Tenth ACM International
  Conference on Future Energy Systems}, Phoenix, AZ, United States, June 2019,
  pp. 24--34.

\bibitem{rahimpour2017non}
A.~Rahimpour, H.~Qi, D.~Fugate, and T.~Kuruganti, ``Non-intrusive energy
  disaggregation using non-negative matrix factorization with sum-to-k
  constraint,'' \emph{IEEE Transactions on Power Systems}, vol.~32, no.~6, pp.
  4430--4441, 2017.

\bibitem{batra2018transferring}
N.~Batra, Y.~Jia, H.~Wang, and K.~Whitehouse, ``Transferring decomposed tensors
  for scalable energy breakdown across regions,'' in \emph{Proceddings of the
  32{nd} AAAI Conference on Artificial Intelligence}, New Orleans, Louisiana,
  United States, Feb. 2018, pp. 240--247.

\bibitem{zamzam2020grate}
A.~S. Zamzam, B.~Yang, and N.~D. Sidiropoulos, ``{GRATE}: Granular recovery of
  aggregated tensor data by example,'' \emph{arXiv preprint arXiv:2003.12666},
  2020.

\bibitem{kelly2015neural}
J.~Kelly and W.~Knottenbelt, ``Neural nilm: Deep neural networks applied to
  energy disaggregation,'' in \emph{Proceedings of the 2nd ACM International
  Conference on Embedded Systems for Energy-Efficient Built Environments},
  Seoul, South Korea, Nov. 2015, pp. 55--64.

\bibitem{zhang2018sequence}
C.~Zhang, M.~Zhong, Z.~Wang, N.~Goddard, and C.~Sutton, ``Sequence-to-point
  learning with neural networks for non-intrusive load monitoring,'' in
  \emph{Proceddings of the 32{nd} AAAI conference on artificial intelligence},
  New Orleans, Louisiana, United States, Feb. 2018.

\bibitem{ouricassp}
F.~M. Almutairi, A.~Konar, and N.~D. Sidiropoulos, ``Scalable energy
  disaggregation via successive submodular approximation,'' in
  \emph{Proceedings of the International Conference on Acoustics, Speech and
  Signal Processing (ICASSP)}, Calgary, AB, Canada, Apr. 2018, pp. 2676--2680.

\bibitem{fujishige2005submodular}
S.~Fujishige, \emph{Submodular functions and optimization}, 2nd~ed., ser.
  Annals of Discrete Mathematics.\hskip 1em plus 0.5em minus 0.4em\relax
  Elsevier, 2005, vol.~58.

\bibitem{bach2013learning}
F.~Bach, ``Learning with submodular functions: A convex optimization
  perspective,'' \emph{Foundations and Trends in Machine Learning}, vol.~6, no.
  2-3, pp. 145--373, 2013.

\bibitem{elnozahy2013comprehensive}
M.~S. ElNozahy and M.~M. Salama, ``A comprehensive study of the impacts of
  {PHEV}s on residential distribution networks,'' \emph{IEEE Trans. Sustainable
  Energy}, vol.~5, no.~1, pp. 332--342, 2013.

\bibitem{wildi2002electrical}
T.~Wildi, ``Electrical machines, drives, and power systems,'' \emph{New Jersey:
  Upper Saddle River}, 2002.

\bibitem{narasimhan2012submodular}
M.~Narasimhan and J.~A. Bilmes, ``A submodular-supermodular procedure with
  applications to discriminative structure learning,'' in \emph{Proceedings of
  the 21st Conference on Uncertainty in Artificial Intelligence}, Edinburgh,
  Scotland, Jul. 2005, pp. 404--412.

\bibitem{iyer2012algorithms}
R.~Iyer and J.~Bilmes, ``Algorithms for approximate minimization of the
  difference between submodular functions, with applications,'' in
  \emph{Proceedings of the 28th Conference on Uncertainty in Artificial
  Intelligence}, Catalina Island, CA, Aug. 2012, pp. 407--417.

\bibitem{iyer2013fast}
R.~Iyer, S.~Jegelka, and J.~Bilmes, ``Fast semidifferential-based submodular
  function optimization: Extended version,'' in \emph{Proceedings of the
  International Conference on Machine Learning (ICML)}, Atlanta, GA, United
  States, June 2013.

\bibitem{edmonds1970submodular}
J.~Edmonds, ``Submodular functions, matroids, and certain polyhedra,''
  \emph{Edited by G. Goos, J. Hartmanis, and J. van Leeuwen}, vol.~11, 1970.

\bibitem{Grotschel}
M.~Gr{\"o}tschel, L.~Lov{\'a}sz, and A.~Schrijver, ``The ellipsoid method and
  its consequences in combinatorial optimization,'' \emph{Combinatorica},
  vol.~1, no.~2, pp. 169--197, Jun. 1981.

\bibitem{kolter2011redd}
J.~Z. Kolter and M.~J. Johnson, ``Redd: A public data set for energy
  disaggregation research,'' in \emph{Workshop on Data Mining Applications in
  Sustainability (SustKDD) in the 17th ACM SIGKDD Conference on Knowledge
  Discovery and Data Mining}, San Diego, CA, United States, Aug. 2011, pp.
  59--62.

\bibitem{beckel2014nilm}
C.~Beckel, W.~Kleiminger, R.~Cicchetti, T.~Staake, and S.~Santini, ``The eco
  data set and the performance of non-intrusive load monitoring algorithms,''
  in \emph{Proceedings of the 1st ACM International Conference on Embedded
  Systems for Energy-Efficient Buildings (BuildSys 2014)}, Memphis, TN, United
  States, Nov. 2014, pp. 80--89.

\bibitem{batra2019towards}
N.~Batra, R.~Kukunuri, A.~Pandey, R.~Malakar, R.~Kumar, O.~Krystalakos,
  M.~Zhong, P.~Meira, and O.~Parson, ``Towards reproducible state-of-the-art
  energy disaggregation,'' in \emph{Proceedings of the 6th ACM International
  Conference on Systems for Energy-Efficient Buildings, Cities, and
  Transportation}, New York, NY, United States, Nov. 2019, pp. 193--202.

\end{thebibliography}

\end{document}